\documentclass[useAMS,usenatbib,twocolumn]{mn2e}

\usepackage{textcomp,graphicx,amsmath,pdflscape,amssymb,setspace,subfigure}

\newcommand\apj{{ApJ}}%
\newcommand\apjl{{ApJ}}%
\newcommand\apss{{Ap\&SS}}%
\newcommand\aap{{A\&A}}%
\newcommand\actaa{{Acta Astronomica}}%
\newcommand\mnras{{MNRAS}}%
\newcommand\pasp{{PASP}}%
\newcommand\nat{{Nature}}%
\title[Generation-II microlensing simulation]
{Second-generation microlensing planet surveys: a realistic simulation}

\author[Shvartzvald \& Maoz]
{Yossi~Shvartzvald$^1$\thanks{E-mail: yossi@wise.tau.ac.il} and
Dan~Maoz$^1$\\
$^1$School of Physics and Astronomy, Tel-Aviv University, Tel-Aviv 69978, Israel\\
}

\date{Accepted 2011 October 14;  Received 2011 October 10; in original form 2011 July 28}

\begin{document}

\maketitle


\setstretch{1}

\begin{abstract}
 \noindent
Microlensing surveys, which have discovered about a dozen extrasolar planets to date, have focused on the small minority of high-magnification lensing events, which have a high sensitivity to planet detection. In contrast, second-generation experiments, of the type that has recently begun, monitor continuously also the majority of low-magnification events. We  carry out a realistic numerical simulation of such experiments. We simulate scaled, solar-like, eight-planet systems, studying a variety of physical parameters (planet frequency, scaling of the snowline with stellar mass $R_{\rm{snow}}\propto M^s$), and folding in the various observational parameters (cadence, experiment duration), with sampling sequences and photometric error distributions taken from the real ongoing experiment.
We quantify the dependence of detected planet yield on cadence and experiment duration, e.g., the yield is doubled when going from 3-hour to 15-minute baseline cadences, or from an 80-day-long to a 150-day-long experiment.
There is a degeneracy between the snowline scaling index $s$ and the abundance of planetary systems that can be inferred from the experiment, in the context of our scaled solar-analog model. After 4 years, the ongoing second-generation experiment will discover of the order of 50 planets, and thus will be able to determine the frequency of snowline planet occurrence to a precision of 10-30\%, assuming the fraction of stars hosting such planets is between 1/3 and 1/10, and a snowline index in the range $s=0.5$ to 2. If most planetary systems are solar analogs, over 65\% of the detected planets will be \textquoteleft Jupiters\textquoteright, five in six of the detected anomalies will be due to a single planet, and one in six will reveal two planets in a single lensing event.
\end{abstract}



\section{INTRODUCTION}
\label{sec:intro}

Among the techniques currently used to discover extrasolar planets, microlensing has some unique capabilities (see, e.g. \citealt{Gould.2010.B}). Apart from the fact that microlensing probes for planets around stars over a large volume of the Galaxy, among both the bulge and the disk population, and can also discover free-floating planets (\citealt{Sumi.2011.A}), it is sensitive to planets at projected host separations that are distinct from those probed by other methods, namely about 1 to 10 AU. This corresponds to the area beyond the \textquotedblleft snowline\textquotedblright\ where gas and ice giants are likely to form. Microlensing is thus complementary to other methods, such as radial velocity or transit surveys, which are most sensitive to planets in close orbits around their host stars, and to direct imaging of planets, which probes larger separations
(e.g. \citealt{Crepp.2011.A}).
To date, however, only about a dozen planets have been discovered via microlensing (\citealt{Bond.2004.A}, \citealt{Udalski.2005.A},  \citealt{Beaulieu.2006.A}, \citealt{Gould.2006.A}, \citealt{Gaudi.2008.A}, \citealt{Bennett.2008.A}, \citealt{Dong.2009.A}, \citealt{Sumi.2010.A}, \citealt{Janczak.2010.A}, \citealt{Miyake.2011.A}, \citealt{Batista.2011.A}), out of the total $\sim$550 known planets\footnote{\tt http://exoplanet.eu}, and the $\sim$1200 candidates recently announced by the Kepler mission (\citealt{Borucki.2011.A}).

The paucity of microlensing-discovered planets is a consequence of the surveying strategy, largely followed to date, which has focused on high-magnification lensing events. As summarized in \citet{Gould.2010.B}, all currently known microlensing-discovered planets were first discovered by one or both of two wide-field, low-cadence microlensing surveys, OGLE (\citealt{Udalski.2009.A}) and MOA (\citealt{Sako.2008.A}). Events whose initial light curves suggested high peak magnifications triggered global networks of telescopes for photometric (and sometimes spectroscopic) follow-up of these specific events.
Indeed, among the 12 known microlensing planets (11 systems, as OGLE-06-109L has two planets) eight were discovered in events with amplifications greater than 100, and three had medium magnifications (greater than 10). All of these events had long event time scales (40-60 days), whereas the typical event time scale is $\sim$20 days (see Section \ref{sec:sim_samp}, below), and high cadences were therefore not essential. Only one event, OGLE-05-169L (\citealt{Beaulieu.2006.A}), was a low-magnification event, with an amplification of 3.0 and a relatively short event time scale of 11 days. The source, however, was a bright star with a baseline magnitude of $I=14.3$, which permitted triggering of the follow-up network nonetheless. All of the above events had peak $I$-band magnitudes brighter than 16.7.

Given the limited observational capabilities to date, the high-magnification follow-up strategy has been the optimal choice. Bright events can be followed up with high cadence using small telescopes, and these events are fortuitously also those with the highest sensitivity to planets. However, only 1\% of all microlensing events have magnifications $>100$. Since the surveys of the Galactic bulge have discovered roughly 700 events per year in recent years, and apparently only a fraction of all stars host planets near their snowlines, only a couple of planets per year have been discovered, resulting in the current microlensing planet tally.

Microlensing survey statistics have been used in the past to constrain the frequency and properties of snowline-region planets by \citet{Albrow.2001.A}, \citet{Gaudi.2002.A}, and \citet{Tsapras.2003.A}. More recently, \citet{Sumi.2010.A} derived the mass-ratio distributions of cold exoplanets, and found that Neptune-mass planets are three times more common than Jupiters beyond the snowline. \citet{Gould.2010.B} find from the high-magnification events that the frequency of Solar-like systems around lens stars is about 1/6. However, two main problems complicate these estimates. One is the small-number statistics they are based on. The other is the complex social process that determines which events are followed up, and how, a process which then needs to be modeled in order to estimate the intrinsic planet properties. \citet{Gould.2010.B} argue that the very chaotic nature of the process results, in the end, in a randomization that minimizes selection effects and gives a high level of completeness. Nevertheless, a more controlled microlensing experiment would provide confidence that this is indeed the case.

It has been clear for some time that, to make progress and overcome these problems, a \textquotedblleft second generation\textquotedblright\ (hereafter generation-II) microlensing survey (e.g., \citealt{Gaudi.2009.A}) must be initiated, one in which a large fraction of all ongoing microlensing events toward the Galactic bulge -- both low and high magnification -- are monitored continuously with a cadence that is high enough to detect planetary perturbations in the light curves. Although only a fraction of low-magnification events will reveal planets, even if they are present around the lens star, the numbers of such events are much larger than high-magnification events. In the balance, such a survey, over several years, could yield a sample of planets that is competitive with those from other techniques. The requirements are straightforward -- a network of 1m-class telescopes situated so as to allow 24-hour coverage of the bulge, equipped with degree-scale imagers.
In June-July 2010, we carried out a 6-week pilot version of such a generation-II experiment. Starting April 2011, we have begun a full generation-II experiment. In this paper, we perform numerical simulations of the possible outcomes of this and similar experiments.

We begin our study by considering a variety of physical possibilities for the abundance of planetary systems and their properties. We then ray trace through the systems to generate a library of microlensing light curves with the correct distribution of observed properties, such as peak magnification, peak magnitude, and timescale. The simulated light curves are then \textquotedblleft observed\textquotedblright\ using the real observational parameters of our ongoing experiment, including sampling, weather gaps and photometric noise. Finally, the data are processed through a realistic detection pipeline, to recover the detection efficiency and the expected number of detections for each combination of physical and observational parameters. The results we present here can therefore serve to predict the power of an experiment to distinguish among models, and to optimize the experiment's yield.
We note that, although generation-II experiments will discover many well-characterized planetary events, our focus is on the mere detection of planetary anomalies, even when they may end up not being unambiguously modeled.

Our methodology builds upon, and extends previous studies of the planet detection efficiencies of microlensing surveys (e.g., \citealt{Gaudi.2000.A}; \citealt{Bennett.2002.A}; \citealt{Bennett.2004.A}; \citealt{Yoo.2004.A}; \citealt{Dominik.2010.A}; \citealt{Penny.2011.A}). However, to our knowledge, ours is the first study to combine the following important elements in such a simulation: we consider full 8-planet solar-analog systems; we investigate the effect of the snowline radius of those systems, for a range of scaling dependences on mass; for our simulated light curve generation, we use the observed distributions of lensing event timescales, impact parameters, and peak magnitudes, and the distributions of lens mass, distance, and velocity implied by Galactic models; we use the real sampling patterns and photometric error distributions (including outliers) of the ongoing generation-II experiment that we simulate.
Eventually, we will use the tools we develop here to constrain the actual physical parameter space of exoplanets probed by microlensing.

In Section \ref{sec:gen_2_net}, we present the observational setup of our generation-II network. In Section \ref{sec:sim_samp}, we describe the simulations, starting with the light curve generation process, and up to the detection procedure. In Section \ref{sec:results} we discuss the results we obtain as a function of the various input parameters. A summary of our results is given in Section \ref{sec:summary}.


\section{The generation-II network}
\label{sec:gen_2_net}

\begin{figure}
\includegraphics[width=0.5\textwidth]{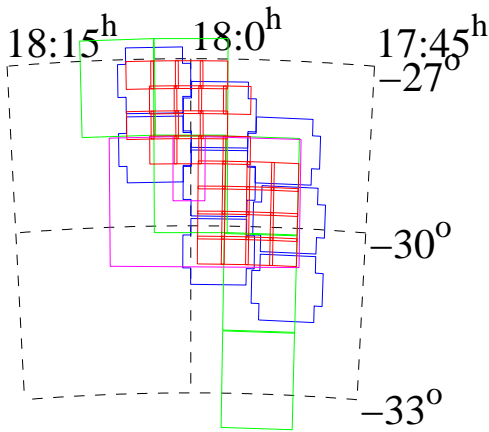}
\caption{ Generation-II network fields: red -- Wise, blue -- OGLE-IV, green -- MOA-II, magenta -- PTF.
\label{fig:fields}}
\end{figure}

The generation-II microlensing experiment is, in essence, a continuous-monitoring, high-cadence survey with a field of view of 8 $\rm deg^2$ centred on the Galactic bulge. The experiment network currently includes four telescopes:
\begin{enumerate}
\item The 1m-telescope at Wise Observatory in Israel with the LAIWO camera, with a 1 $\rm deg^2$ field of view (\citealt{Gorbikov.2010.A}). The cadence for each of the eight Wise fields is $\sim$30 minutes.
\item The Las Campanas Observatory 1.3m Warsaw University telescope in Chile with the OGLE-IV camera, with a 1.4 $\rm deg^2$ field of view (\citealt{Udalski.2009.A}). The OGLE-IV phase has three extremely high-cadence fields observed every $\sim$15 minutes and five high-cadence fields observed once per $\sim$45 minutes.
\item The 1.8m MOA-II telescope in New Zealand with the MOA-cam3 camera, with a 2.2 $\rm deg^2$ field of view (\citealt{Sako.2008.A}). The MOA-II phase has six extremely high-cadence fields (every $\sim$15 minutes) and six high-cadence fields (every $\sim$45 minutes); most of these latter fields are outside the full network footprint.
\item The Palomar Observatory 1.2m Oschin telescope in California with the PTF camera, with a 7.8 $\rm deg^2$ field of view (\citealt{Law.2009.A}) and a cadence of $\sim$40 minutes. This telescope participates only during one month every year, when the network has the highest continuous coverage. Furthermore, the PTF footprint overlaps with only 4 $\rm deg^2$ of the Wise fields.
\end{enumerate}
The network parameters are summarized in Table \ref{table:network}.
Figure \ref{fig:fields} shows the region of the Galactic bulge covered by the network. The 8 $\rm deg^2$ covered by Wise, OGLE, and MOA include the fields with the highest microlensing event rates in previous OGLE and MOA seasons.

\begin{table}
\caption{The generation-II network}
\begin{tabular}{l|c|c|c|c}
\hline
\hline
Group (site) & Aperture & FOV & Cadence & High cadence\\
 & [m] & [$\rm deg^2$] & [min] &  area [$\rm deg^2$]\\
\hline

Wise (Israel) & 1.0 & 1.0 & 30 & 8.0 \\

OGLE (Chile) & 1.3 & 1.4 & 15 (45) & 4.2 (7.0)\\

MOA (NZ) & 1.8 & 2.2 & 15 (45) & 13.2 (13.2)\\

PTF (California) & 1.2 & 7.8 & 40 & 7.8 \\

\hline
\end{tabular}
Note -- numbers in parentheses are the cadences and areas of the lower-cadence OGLE and MOA fields.
\label{table:network}
\end{table}

\section{EXPERIMENT SIMULATION}
\label{sec:sim_samp}

In this section, we describe the various stages of our numerical simulation of the generation-II microlensing planet survey.
\subsection{Source-lens configuration}
\label{subsec:source_lens_conf}
\begin{figure*}
\begin{minipage}{\textwidth}
\center
\begin{tabular}{cc}
\includegraphics[width=0.5\textwidth]{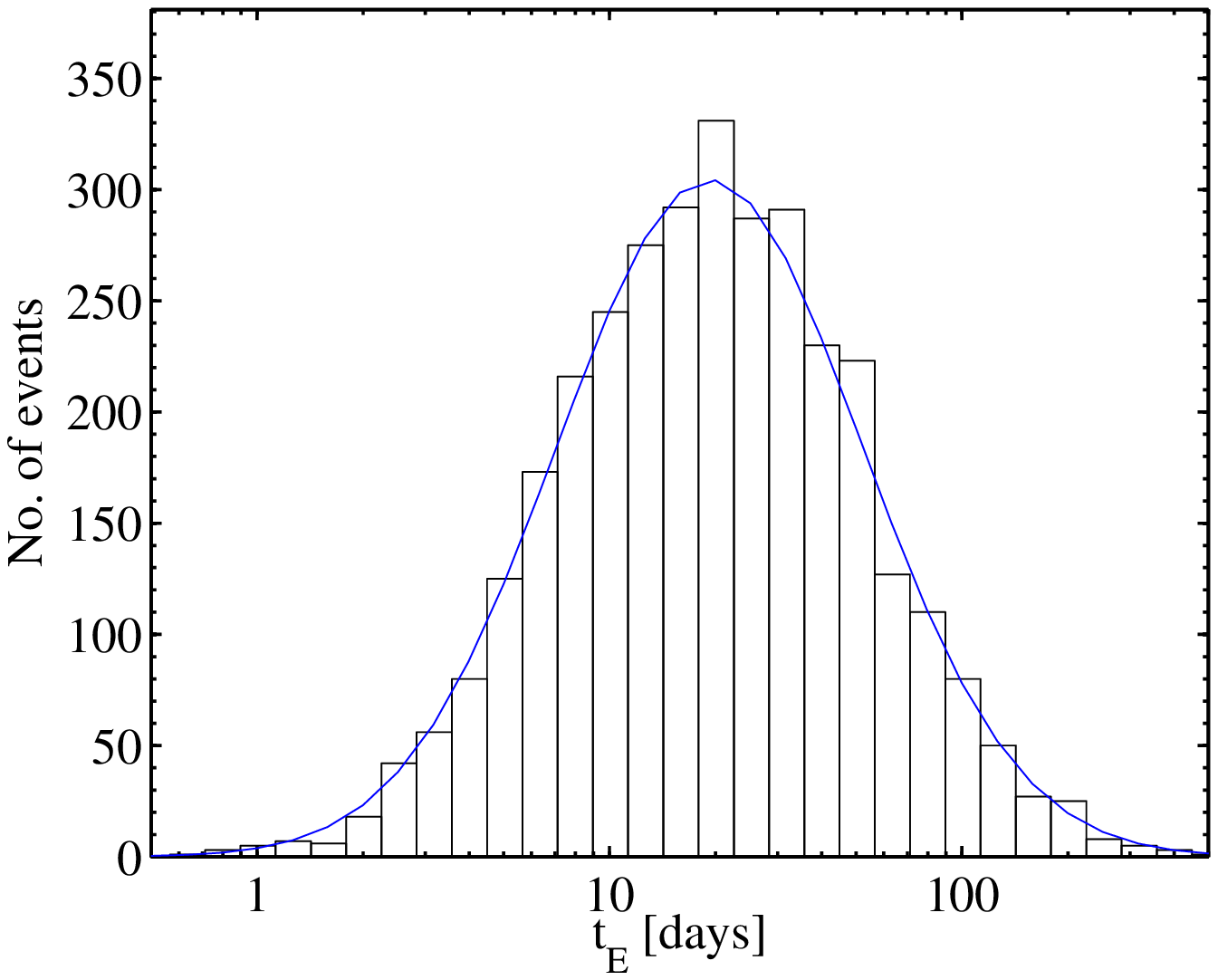} & \includegraphics[width=0.5\textwidth]{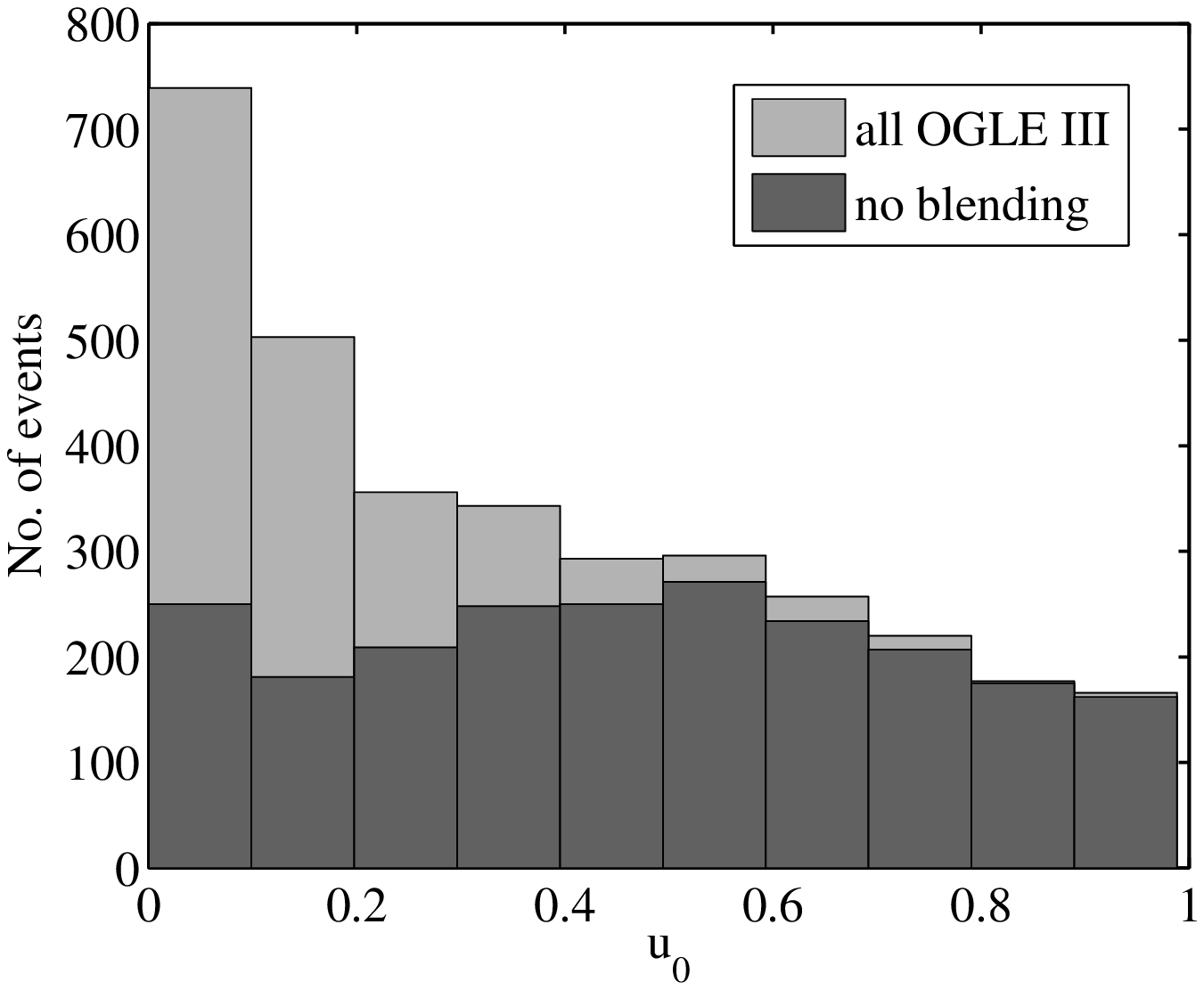}
\end{tabular}
\caption{ {\it Left}: Observed distribution of event time scales, $t_E$, for all OGLE-III ($2002-2009$) events with $u_0\leq1$, obtained from the OGLE database. The solid curve is a lognormal fit, $N\propto{{\rm exp}(-\frac{(\log(t_E)-\mu)^2}{2\sigma^2})}$, with $\mu=1.28$ and $\sigma=0.43$.  {\it Right}: Distribution of impact parameter, $u_0$, for all OGLE-III events with $u_0\leq1$ (light-gray). Due to selection favoring the detection of highly magnified faint events over weakly magnified faint events, there are more events at smaller impact parameters. The dark-gray histogram shows the distribution only for unblended (i.e. relatively bright) events, which are seem to be distributed uniformly, as expected. \label{fig:ogle_t_u}}
\end{minipage}
\end{figure*}

A point-mass microlensing event is characterized by its angular Einstein radius, $\theta_E$:
\begin{equation}
\theta_E = \sqrt{\frac{4GM_L}{c^2}\frac{D_S-D_L}{D_SD_L}}
\end{equation}
where $M_L$ is the mass of the lens star, $D_S$ and $D_L$ are the distances to the source and lens stars, respectively, $G$ is the gravitational constant, and $c$ is the speed of light. The time scale of the event, $t_E$, is the time it takes the source star to cross the angular Einstein radius due to the star's relative transverse velocity, $v_t$. The angular position of the source star at time $t$, neglecting parallax (due to Earth's acceleration), orbital motion of the lens due to a companion, and \textquotedblleft xallarap\textquotedblright\ orbital motion of a binary source is:
\begin{equation}\label{eq:ut}
u(t)=\sqrt{u_0^2+\left(\frac{t-t_0}{t_E}\right)^2}
\end{equation}
where $u(t)$ is the angular separation between the source and the lens in units of $\theta_E$, $u_0$ is the impact parameter -- the minimum angular separation between the source and the lens, and $t_0$ is the time at minimum separation.
Among $t_E$, $u_0$ and $t_0$, $t_E$ contains the interesting physical properties of the source-lens configuration (distances, relative velocity and lens mass). Our first step is to model those physical properties. \citet{Dominik.2006.A} has used Galactic structure models to find the probability functions for the lens mass, lens distance, and transverse velocity, as a function of the event time-scale, $t_E$, in the direction of the Galactic bulge through Baade's window (Galactic longitude and latitude $[l,b] = [1^\circ,-3.9^\circ]$), for a source star at $D_S=8.5$~kpc (see his figure 5). We create our simulated sample of lenses by choosing $t_E$ from the observed distribution of all OGLE-III (2002-2009) events (Fig. \ref{fig:ogle_t_u}), as listed in the OGLE database\footnote{\tt http://ogle.astrouw.edu.pl} and, for each $t_E$, drawing $M_L, D_L$ and $v_t$ from the \citet{Dominik.2006.A} probability functions.

The impact parameter, $u_0$, should, in principle, have a uniform distribution. In practice, however, due to the flux limits of the actual observations, there is a selection effect that, for intrinsically faint sources, favors the detection of high-magnification events (i.e., small $u_0$). In line with our objective of making our simulations as realistic as possible,
we therefore choose also the impact parameter from the distribution of all OGLE-III events (Fig. \ref{fig:ogle_t_u}). We  confirm that, for OGLE events where the source was unblended, implying it was always above the limiting magnitude, the distribution of $u_0$ is consistent with being uniform (up to Poisson fluctuations). With this, we have the full source-lens configuration of an event.

\subsection{Planetary systems}
\label{subsec:ps}

The characteristics and diversity of \textquotedblleft normal\textquotedblright\ extrasolar planetary systems are still virtually unknown.
Although there is an infinite variety of ways in which planetary systems could, in principle, be structured, with planets of all masses at all radii, real multiple planetary systems obey physical constraints, resulting from formation history and stability criteria.
A Copernican approach is therefore to assume as a working hypothesis (to be tested by observations) that the Solar system is a typical planetary system, as hinted by the microlensing discovery of a scaled-down Sun-Jupiter-Saturn system in OGLE-06-109L (\citealt{Gaudi.2008.A}; \citealt{Bennett.2010.A}). Indeed, a major advantage of microlensing as a planet-hunting technique is that it is the most sensitive method available for the detection of the snowline region of such analogs (\citealt{Gould.2006.A}, \citealt{Gould.2009.A}).
We thus choose our simulated planetary systems to mimic the Solar system, with eight coplanar planets in circular orbits, having the same mass and distance ratios as the Solar system. The stellar mass, $M_L$, as noted above, is drawn from the \citet{Dominik.2006.A} probability function.

The location of the snowline at the time of planet formation is thought to set the border between rocky planets and gas/ice giants, and therefore the dimensions of a  planetary system may scale with the snowline radius. The snowline radius itself will depend on the stellar mass, in a way that can be parametrized as a power-law, $R_{\rm{snow}}\propto M^s$. The value of the index $s$ has been debated (e.g., \citealt{Lecar.2006.A}, \citealt{Kennedy.2006.A}, \citealt{Kennedy.2008.A}). For example, considering just the equilibrium temperature around main-sequence stars, $s\approx 2$ would be expected. However, various effects at the time of planet formation in a protoplanetary disk around a pre-main-sequence star could modify this value. In our simulation we explore
the range $0.5\leq s\leq 2$.
The zeropoint for the scaling is Jupiter's orbit in the solar system (5.2 AU).

Finally, for each simulated event, the inclination to our line of sight of the planetary system, the phase of each planet  in its orbit, and the source trajectory angle with respect to the orientation of the orbital plane of lens system on the sky, are chosen at random.

\subsection{Light-curve generation}
\label{subsec:lc_gen}

For a lens that is a star with a planetary system that includes $n-1$ planets, the lens equation that relates the source position, $\overrightarrow{\beta}$, and the image positions, $\overrightarrow{\theta}$, in the lens plane is:
\begin{equation}\label{eq:lens_eq}
\overrightarrow{\beta}=\overrightarrow{\theta}-\theta_{E,L}^{2}\sum^{n}_{i=1}{\frac{q_i}{\overrightarrow{\theta}-\overrightarrow{\theta_i}}}
\end{equation}
where $\theta_{E,L}$ is the angular Einstein radius of the lens star, $q_i$ is the mass ratio between the $i$th planet and the lens star ($q_i=1$ for the lens star) and $\theta_i$ is the position of the $i$th planet. The magnification is the ratio of the summed solid angles subtended by all the images to that of the solid angle subtended by the source.

\begin{figure*}
\begin{minipage}{\textwidth}
\center
\begin{tabular}{c}
\includegraphics[width=0.9\textwidth]{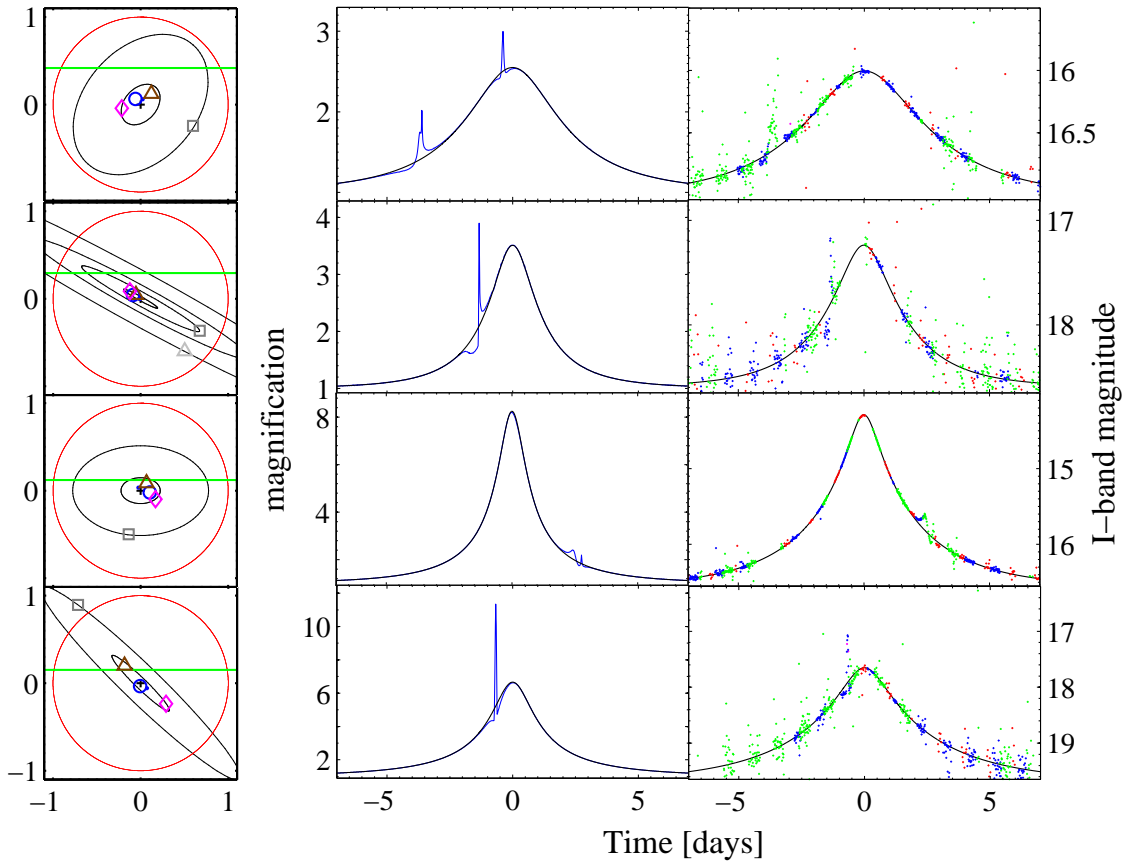}
\end{tabular}
\caption{Examples of simulations with our light-curve generator.
{\it Left column}: Several configurations of the lens plane, in units of the lens star's Einstein radius, of projected Solar-like planetary systems around a lens star with a snowline power-law index of $s=1$. The projected circular orbits of some of the planets are shown (black ellipses). The red circle is the host-star Einstein radius and the symbols mark the projected positions of the 8 planets of the system, some of them outside the area shown (Mercury - blue point, Venus - blue circle, Earth - brown triangle, Mars - pink diamond, Jupiter - gray square, Uranus - light-gray triangle). The green line is the trajectory of the source star and the black cross is the host star. {\it Middle column}: Light curves (blue) obtained by ray tracing from the observer, through the planetary system, to the source. Black curves are for the lens star without planets. {\it Right column}: Light curve for each system, as \textquoteleft observed\textquoteright\ by the four observatories of the generation-II microlensing network (green - MOA, blue - OGLE, red - Wise, magenta - PTF). The black curves are the light curves for the lens star without any planets. \label{fig:sim_example}}
\end{minipage}
\end{figure*}

In principle, solving the lens equation for the image positions, given a source position, requires solving a polynomial of degree $n^2+1$ although some of those solutions are not physical, and hence the number of variables can be reduced to $2n$ (\citealt{Witt.1990.A}). Given the large number of planets per star that we wish to investigate in our simulations, we choose to generate lensed images via inverse ray-shooting (\citealt{Schneider.1988.A}) and thus solve for the magnification $A(t)$ of the source flux. In this brute-force approach, we divide the lens plane onto a grid and use the lens equation directly to map the lens plane onto the source plane. In order to achieve fast performance and high accuracy, we use an adaptive grid that increases the lens plane resolution at the positions of the images. We can thus also account for finite source effects (\citealt{Nemiroff.1994.A}) and for the limb-darkened profile of the source star. For simplicity, in this work we use a constant source size of 1$R_\odot$, and a uniform source surface brightness.
Giant sources with radii of $\sim$ 5--10 $R_\odot$, which can be a substantial fraction of the sources, can eliminate small-deviation detections due to their larger area but, on the other hand, have a larger effective cross-section for planet lensing.
Assuming a rough border in unlensed source magnitude of $I=17.5$ mag between giants and dwarfs in the bulge we find that  about 20\% of current OGLE microlensing events have giant sources. To gauge their effect, we found that, even if all the sources in our simulations, described below, are assumed to be giants, fewer than 5\% of detections are missed.
Figure \ref{fig:sim_example} shows several examples of planetary configurations (left panels) and the microlensing light curves that they generate (central panels).

\subsection{Event sampling}
\label{subsec:ev_sampling}

The last step in our procedure is to simulate realistic observations of the light curves that we have generated. We include the following effects. First, we sample each event in a way that mimics the real observations, including gaps due to weather, full moon, and technical problems.
Specifically, we use the actual sampling patterns of the generation-II network from the pilot campaign.
The cadence of the sampling is crucial for planet detection because the time scale for the duration of the planetary deviation is relatively short, $t_{\rm{planet}}\approx\sqrt{q}\cdot t_E$ (\citealt{Gould.1992.A}). A baseline magnitude needs to be assigned to the event, which may be below the detection limit of a given telescope. Frequently, the source is blended with other stars close to the line of sight. Photometric errors added to each measurement need to include Poisson noise from both the source and the stellar backgrounds. In a real experiment, however, difference-image analysis (DIA) is usually performed in order to achieve accurate photometry in crowded fields. The photometric errors on the difference residuals are often under-estimated and there may be residuals due to imperfect subtractions, resulting in outlier points in the light curves, and these errors need to be simulated as well.

We include all of these effects as follows. Every simulated event is matched to a real event, drawn from the list of all past OGLE-III events. For each such event, as noted in Section \ref{subsec:source_lens_conf}, we use the actual $t_E$, and also the source base magnitude and blending fraction. Each light curve is sampled with a pattern of \textquotedblleft minimal\textquotedblright\ cadences (from 15 minutes up to 3 hours) that are later corrupted by the actual sampling
patterns of the 2010 pilot experiment, including measurements lost to weather, full moon, and so on.
Furthermore, our sampling reflects the different sampling patterns of each observatory. Thus, for example, changing the minimal cadence of a simulated run from 15 minutes to 30 minutes does not affect the PTF sampling pattern, which already has a 
40-minute cadence to begin with.
The total duration of the experiment per bulge observing season is chosen from the range 30--150 days. For each observatory, we construct a realistic measurement-scatter distribution as a function of apparent magnitude (including the blended light contribution). For OGLE and MOA, we used representative events from the OGLE-III 2009 season (the last season of OGLE III), and from the MOA-II 2010 season (obtained from the MOA database\footnote{\tt http://www.phys.canterbury.ac.nz/moa}), respectively. The scatter was obtained from the differences between the observed photometry and the best point-source-point-lens fit to each event. For Wise and PTF, we used the MOA-II scatter which, based on the pilot results, is similar for all three observatories. This procedure naturally accounts for all the uncertainties that arise from the observations and the data reductions, including outlier photometric measurements (usually due to DIA residuals).
Figure \ref{fig:sim_example} (right panels) shows four examples of final, sampled and noised, light curves.

\subsection{Anomaly detection}
\label{subsec:anomaly_det}

The most general solution of a microlensing event that has an anomaly caused by just one planet requires fitting for over a dozen parameters, or even more in the case of multiple planets. These parameters represent the source-lens configuration, the planetary system, the dynamics of the system (i.e, parallax, orbital motion, xallarap), the source size, and the limb-darkening profile. There are degeneracies between some of these parameters that can sometimes cause poor characterization of the system or even the false detection of a planet.

However, in order to estimate the abundance of planetary systems, we do not need to fully characterise every event, but only to find the signature of a planetary perturbation with respect to a point-lens light curve. Since the mass ratio between a planet and its host star, $q_i$, is relatively small, the deviation will generally be localized in time, when the source passes close to, or crosses, caustics (central or planetary), while the rest of the light curve will stay mostly unperturbed (more precisely, the time scale of the full light curve will change with respect to a planet-less system, especially for higher-mass ratios, but the form of most of the light curve remains that of a point lens). After fitting a global point-lens model, we therefore search each simulated light curve for a localized deviation from that model, using a local $\chi^2$ test.

Since a $\chi^2$ test requires reliable error estimates, we first obtain an empirical estimate of the photometric errors, from the scatter of the measurements around the point-lens model for each light curve of every observatory's dataset, as a function of event magnitude. However, when we previously noised our simulated light curves (Section \ref{subsec:ev_sampling}, above), we included also the outlier population of measurement errors that exists in the real data. Using an iterative $\sigma$-clipping  procedure (i.e., we repeatedly fit a Gaussian to the error distribution, and reject 2$\sigma$ outliers), we now eliminate these outliers and find the best--fitting Gaussian error distribution (irrespective of the reported errors, which are often underestimated). With our derived errors, the $\chi^2$ of the entire light curve will naturally be close to the number of degrees of freedom. However, when applied piecewise to sections of the light curve, $\chi^2$ can reveal the presence of a planet via the temporal condensations of \textquoteleft outliers\textquoteright\ that the planetary deviation causes.
We define a ``running'' local $\chi^2$ estimator:
\begin{equation}
\chi^2_{\rm local} = \sum\limits_{i=1}^{31} \dfrac{(f_i-f_{\rm pl})^2}{\sigma_i^2}
\end{equation}
where $f_i$ and $\sigma_i$ are the ``observed'' flux and the derived error of each epoch, respectively, and $f_{\rm pl}$ is the point-lens model flux at that time.
The local $\chi^2$ is repeatedly calculated by advancing one observed epoch at a time.
We empirically found the 31-consecutive point length to be optimal
for detecting short-period deviations without increasing false--positive detection due to random error correlations.
An anomaly is considered detected if the $\chi^2$ probability is smaller than 0.3\%
($\chi^2 >53.3$ for 28 degrees of freedom -- 31 points minus three fitted parameters: $t_E$, $u_0$ and $t_0$).
In a real experiment, the method provides also a second-order calibration between observatories, if the global photometric scatter (for the entire light curve) of a specific observatory is not symmetric around zero.

Photometric outlier points that survived the previous filtering will affect the local $\chi^2$ test and lead to false detections. In order to avoid this, our anomaly detection criterion also requires at least three consecutive data points, each with a 3$\sigma$ deviation from the point-lens model
(as in the OGLE Early-Early Warning System anomaly detector, \citealt{Udalski.2003.A}).

Naturally, a more sophisticated or flexible detection criterion could be devised, one that adapts to different
cadences, weather gaps, and deviation timescales. However, for our present purposes, the main point is that, as long
as the same detection criteria are used on the real and the simulated data, the statistical properties of planetary systems can still be correctly estimated.

\subsection{Monte Carlo simulation}
For every combination of physical and observational parameters that we study, we generate 3333 simulated light curves, as described above, pass them through our detection filter, and note the number of planet detections. Naturally, if only a fraction $f$ of lens stars host planetary systems, the number of detections will be reduced correspondingly, by a factor $f$. We examine combinations of three input parameters in each set of simulation: $s$ -- the snowline scaling power-law index; $\tau_{\rm obs}$ -- the total observing period, from 1 month (similar to the generation-II pilot campaign of 2010) and up to 150 days, which is approximately the total visibility period of the Galactic bulge every year; and the minimal cadence, from 15 minutes, as in the high-cadence fields of OGLE and MOA, to 3 hours.

We recall that the minimal cadence is just the baseline for the real observational sequence of the simulation, which uses the actual generation-II pilot campaign observing times.
Similarly, the total observing period includes the real gaps (due to, e.g. weather, full moon, technical downtime) of the observations.

The number of expected detections in a generation-II seasonal campaign is simply the number of events times the detection fraction, and times the true planet abundance, $f$. The number of microlensing events per season in the Galactic bulge detected by OGLE-III and MOA-II has been about 600-650. With OGLE-IV, the event rate will exceed 1000 events per season. The fraction of ongoing events inside the 8 $\rm deg^2$ footprint that was continuously monitored by the generation-II pilot has been $34\%$, so, in what follows, the number of expected planetary detections assumes 340 events per season. The actual number of detections in a real experiment will be distributed according to Poisson statistics around the expectation values that we predict.

\subsection{Binary contamination}
A potentially important contaminant of a microlensing planet survey is \textquotedblleft false positives\textquotedblright\ due to binary, as opposed to planetary, systems. With well-sampled, high S/N data, binary and planetary deviations can generally be distinguished because of the different secondary/primary mass ratios. As already noted, the ratio between the timescales of the anomaly and the full event is of order the square root of the mass ratio. However, in a real generation-II experiment, the binary or planetary nature of some events may remain ambiguous. To quantify this source of systematic error, we simulate a population of binary lenses. We note that, since the mass ratio is no longer very small in the case of binaries, the perturbation due to the companion mass is no longer local and the single-lens model that best fits the data, before applying the detection criteria, will generally be different from that of a single lens with the mass of the primary.

\begin{figure}
\includegraphics[width=0.5\textwidth]{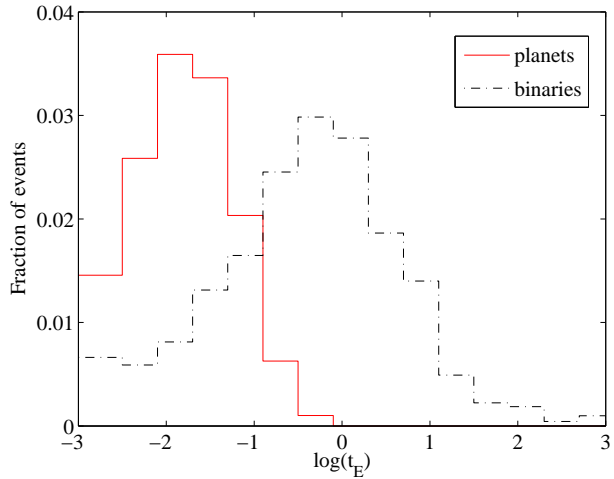}
\caption{ Event distribution as a function of total duration of a 3$\sigma$ deviation from a point-lens model, in units of the Einstein-radius crossing time, for binaries (black) and for planets (red). Planetary deviations are typically much shorter than those due to binaries. \label{fig:binary}}
\end{figure}

For the primary-mass stars in the simulated binaries, we use OGLE-III events as before, and the same mass, velocity and distance distributions. We assume that the physical binary separation, $a$, is distributed as $f(a)\sim a^{-1}$ (\citealt{Opik.1924.A}) although other forms can be considered (e.g. \citealt{Duquennoy.1991.A}). The mass ratio $q$ of the binary is drawn from a flat distribution, $f(q)={\rm const}$, which appears to be generic to a wide range of binary masses and separations (e.g. \citealt{Soderhjelm.2000.A}, \citealt{Mazeh.2003.A}). The binary contamination in a microlensing survey will naturally depend both on the fraction of stars that are in binaries and on the fraction of stars that host planetary systems. \citet{Lada.2006.A} has summarized several studies on binarity fraction. He emphasizes that stellar multiplicity is a rising function of the stellar mass, and therefore most stars are single (since low-mass stars are the most common ones). For example, the single-star fraction is $\sim$70\% among M-type stars and $\sim$45\% among G-type stars.

\begin{figure*}
\begin{minipage}{\textwidth}
\center
\begin{tabular}{cc}
\includegraphics[width=0.5\textwidth]{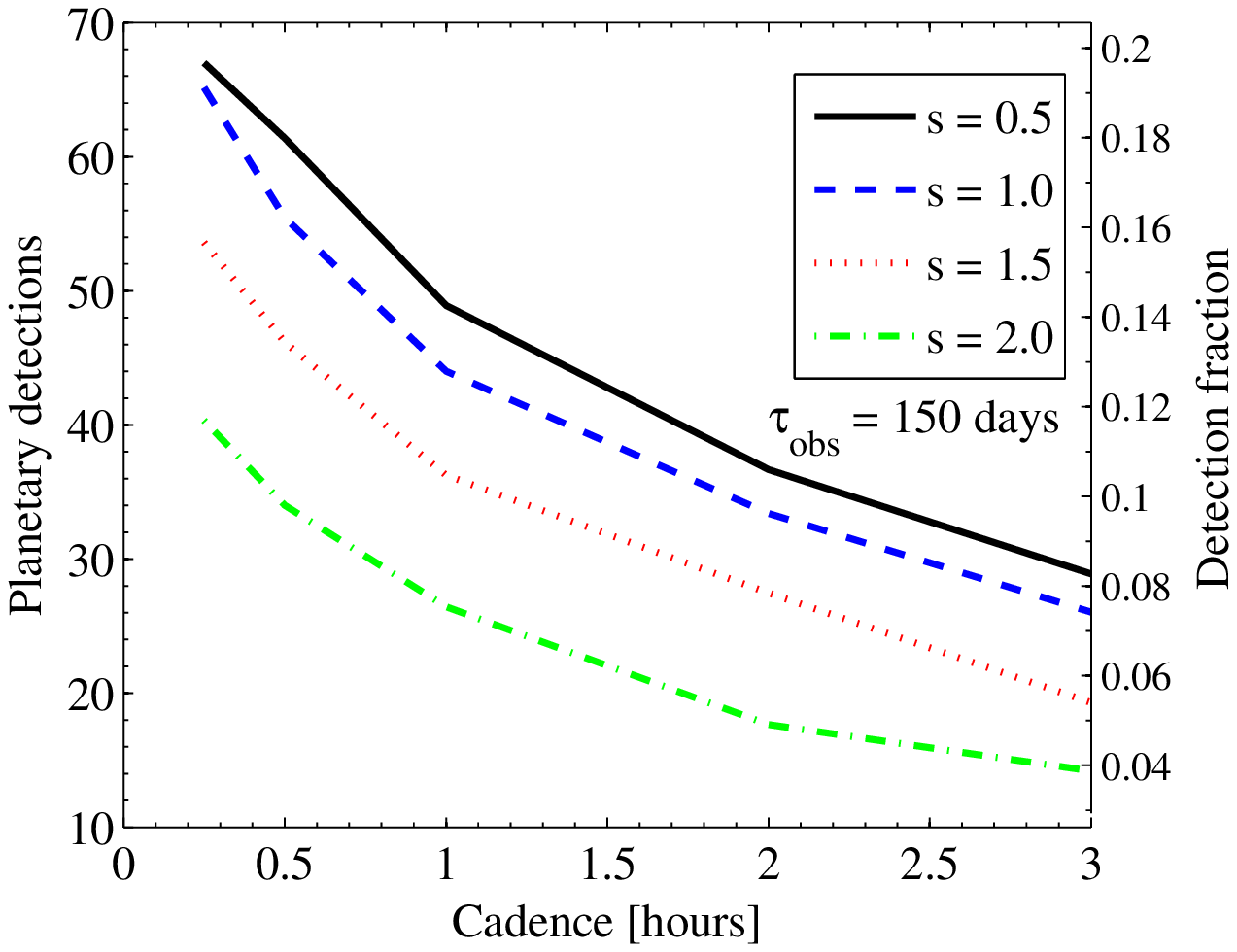} & \includegraphics[width=0.5\textwidth]{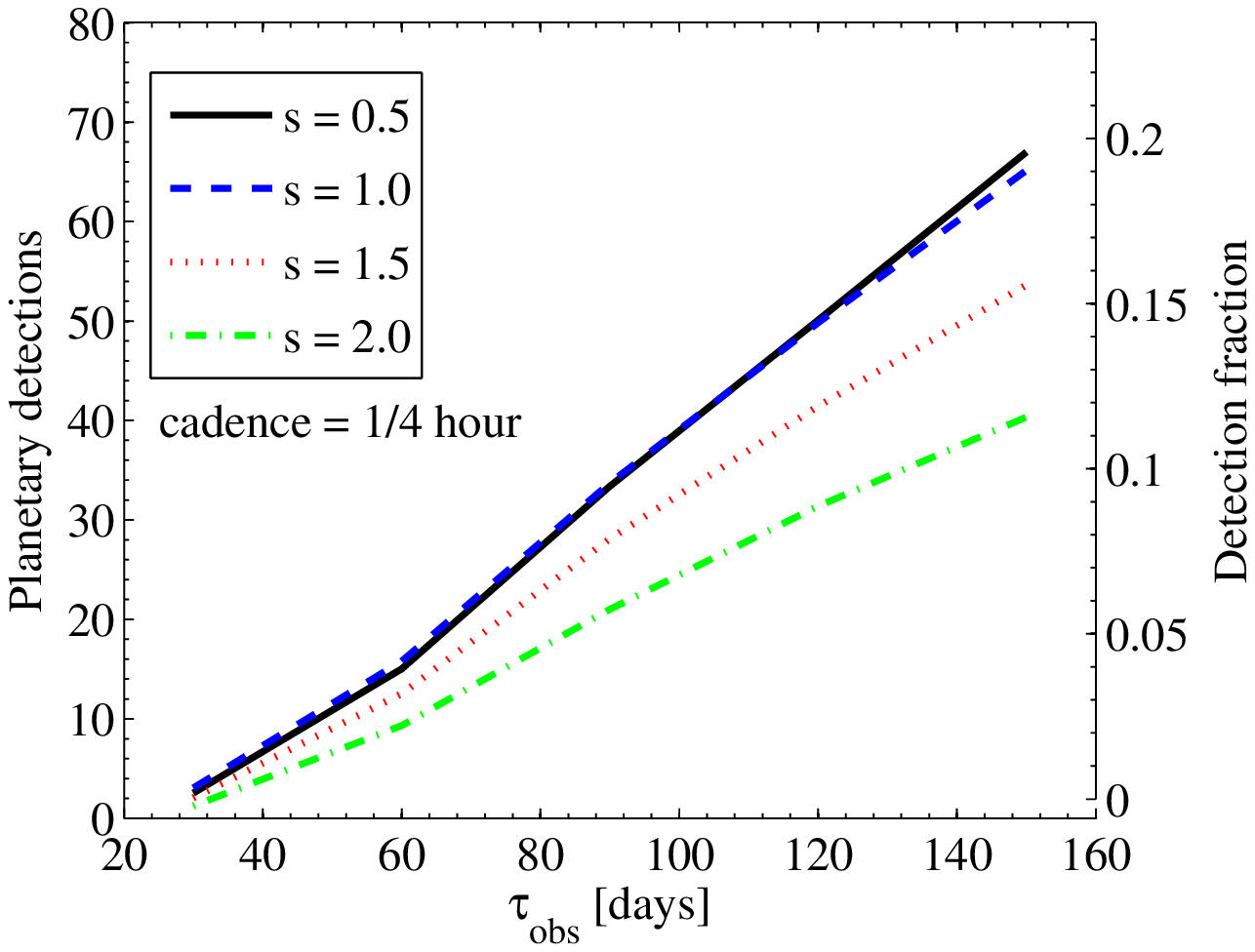}
\end{tabular}
\caption{ {\it Left}: Number of planetary detections as a function of the minimal cadence for different snowline indices $s$, for a total observing period of 150 days, with 340 events per season in the generation-II survey footprint, and assuming each lens star hosts a solar-analog planetary system.
Multiple-planet detections are counted as one event.
The right-hand vertical axis is the detection fraction among all of the events monitored. For a fraction $f$ of lens stars hosting planets, the curves will scale down correspondingly. {\it Right}: Planetary detections as a function of the total observing time per season, $\tau_{\rm obs}$, with a minimal cadence of 15 minutes for different snowline indices $s$, as before, assuming each lens star hosts a planetary system. \label{fig:cadence_days}}
\end{minipage}
\end{figure*}

Figure \ref{fig:binary} shows, for the simulated populations of planets and binaries, the distributions of total detected deviation time (i.e. the total time there was a 3$\sigma$ deviation from the best-fit point-lens light curve). The deviation time is in units of $t_E$ for each event. As expected, the typical binary deviations are of order $t_E$ while the most common planetary deviations are two orders of magnitude more brief. However, there is an overlap in the distributions, which can be a source of false positive planet detections. For example, assuming a planetary frequency of $f=1/6$, as estimated by \citet{Gould.2010.B}, and the \citet{Lada.2006.A} binarity fraction as a function of primary mass, the contamination of binaries to the short deviation events (total deviation less than $0.1t_E$) would be 33\%.

The binary population is currently much better characterised than the planetary one. Since binaries completely dominate
the events with perturbations having timescales $>0.1 t_E$, that part of the distribution can be used to anchor a model for the binary population, and thus to estimate and statistically subtract off the binary contamination on short timescales. Furthermore, although our emphasis has been on the detection of deviations from single-lens light curves, detailed modeling that discriminates between planets and binaries will certainly be possible for a fraction of the events. Such modeling will be possible either with the Generation-II data themselves, or with the addition of followup data from networks that focus on anomalous events, such as MicroFUN (\citealt{Gould.2010.B}) and PLANET (\citealt{Albrow.1998.A}).

\section{RESULTS}
\label{sec:results}

In this section, we present the results of our simulations for various observational and physical input parameters.

\subsection{Observational parameters}
\label{subsec:obs_pars}

\begin{figure*}
\begin{minipage}{\textwidth}
\center
\begin{tabular}{cc}
\includegraphics[width=0.5\textwidth]{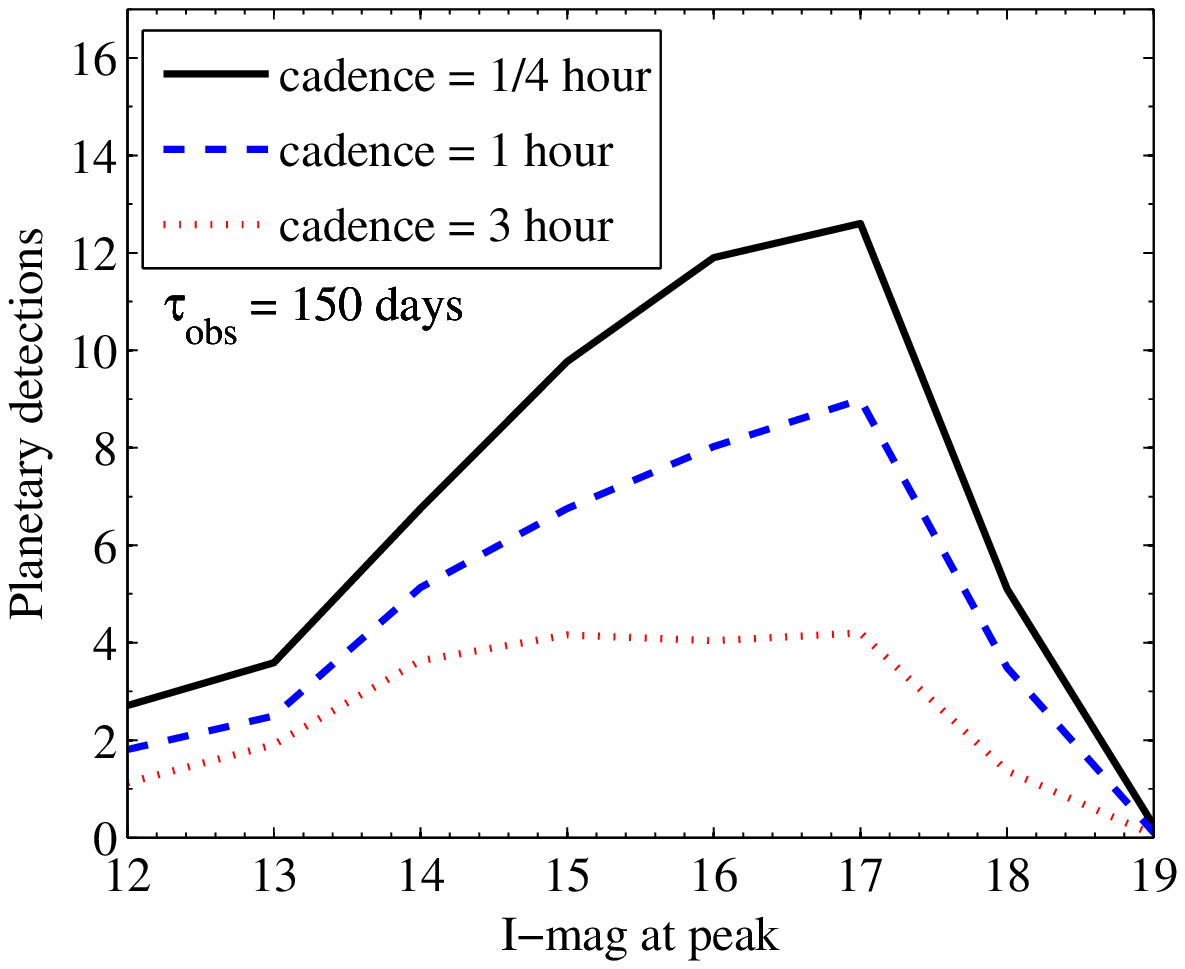} & \includegraphics[width=0.5\textwidth]{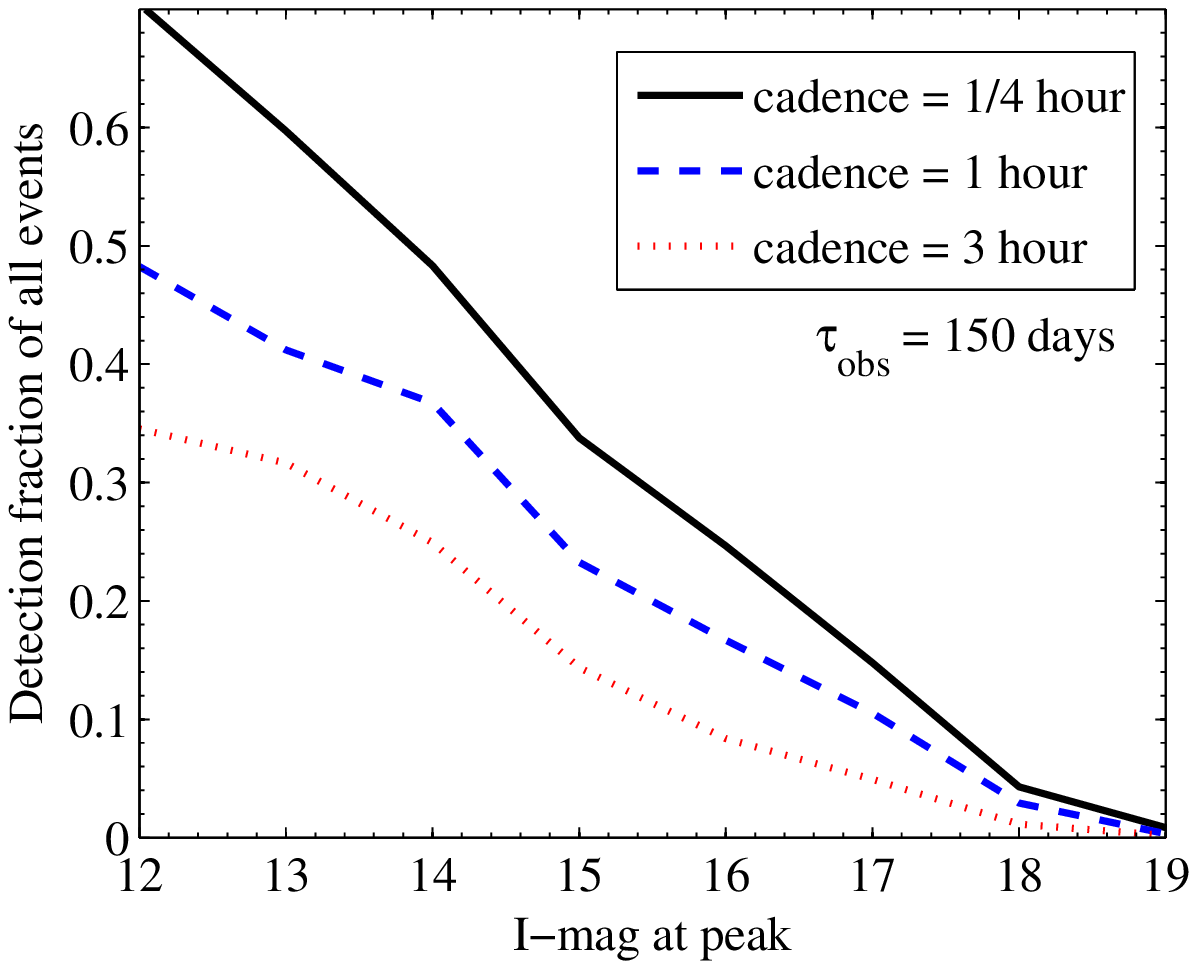}
\end{tabular}
\caption{ {\it Left}: Number of detections as a function of peak $I$-band magnitude of an event, for a 150-day season and various cadences, with results averaged over all snowline indices. The maximum is at $\sim$17 mag. {\it Right}: Detected fraction of events versus peak $I$-band magnitude. The increase towards brighter events results both from the more accurate photometry and from the higher sensitivity to planets of high-magnification events. \label{fig:maxI}}
\end{minipage}
\end{figure*}

Figure \ref{fig:cadence_days} shows the total number of planetary detections per season as a function of the minimal observing cadence, assuming every lens star hosts a planetary system, a total observing period of 150 days and for a range of snowline power-law indices, $s$. We find that over half of the detections are missed by going from a minimal cadence of 15 minutes to 3 hours. This is simply a consequence of the loss of low-mass planet detections, with their short timescales.
Figure \ref{fig:cadence_days} also shows the total number of planetary detections as a function of the total observing period, assuming each lens star hosts a planetary system, a minimal cadence of 15 minutes and the various snowline indices. As expected, the number of detections rises, more or less linearly with the total observing time, because the total number of events is proportional to the total observing period. In addition, there is a second-order contribution to the detected number for long experiment durations because a detectable deviation in long events (long $t_E$) can be missed by a brief campaign.

The detection fraction is naturally sensitive to the precision of the photometry, which depends on the event magnitude. As mentioned in Section \ref{sec:intro}, this is already evidenced in the planets that have been discovered to date via microlensing, which have typical peak event magnitudes of $\sim$15. These events were either of high-magnification, or with sources that were bright in the first place, and with long event time scales, $t_E$. Generation-II experiments improve upon this limitation with high cadences and accurate photometry.
Figure \ref{fig:maxI} shows that the number of planetary detections as a function of $I$-band magnitude at peak,
for the type of events that are detected by the current surveys (OGLE and MOA),
has a maximum at magnitude $\sim$17 for a 150-day season and various cadences, averaged over all snowline indices.
The peak results from, on the one hand, the increased number of events with low magnifications and faint baselines, and on the other hand, the reduced sensitivity of low-magnification and faint events.
The detected fraction of events per peak $I$-band magnitude increases about linearly with magnitude for brighter events (Figure \ref{fig:maxI}, right panel). This folds in both the high photometric precision of the bright events and the higher sensitivity to planets of high-magnification events, which we further explore in the next section.

\subsection{Physical parameters}
\label{subsec:physical_constrains}

\begin{figure}
\includegraphics[width=0.5\textwidth]{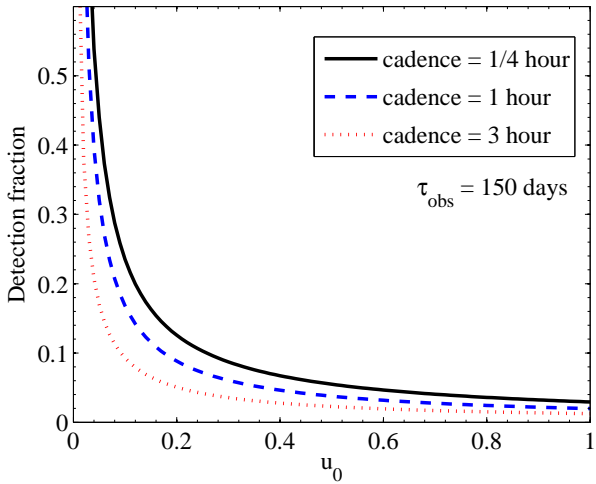}
\caption{ Planet detection efficiency as a function of impact parameter, for a 150-day season and various cadences, with results averaged over all snowline indices. \label{fig:u_0}}
\end{figure}

The sensitivity to planets in high-magnification events is close to 100\% for planets projected near the Einstein radius, because of their signatures on the central caustic (\citealt{Griest.1998.A}).
The sensitivity naturally declines for lower impact parameters. Our simulations permit us to study this question for the specifics of the generation-II experiment.
Figure \ref{fig:u_0} shows the detection fraction of events as a function of the impact parameter $u_0$ (with results averaged over all snowline indices).
The detected fraction depends on impact parameter approximately as $u_0^{-1}$.
Among intermediate magnification events, with impact parameters $0.1 < u_0 < 0.3$ (corresponding to amplifications $10 > A_{\rm max} > 3.4$) planetary systems are detected with $\sim$17\% efficiency for the optimal observing setup (150-day season and minimal cadence of 15 minutes).

\begin{figure}
\includegraphics[width=0.5\textwidth]{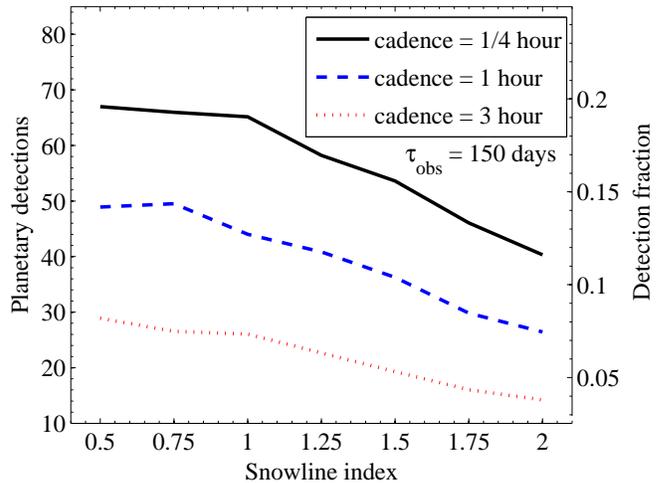}
\caption{ Planetary detections as a function of the snowline scaling index $s$, $R_{\rm{snow}}\propto M^s$, for various cadences. As in Figs. \ref{fig:cadence_days}-\ref{fig:maxI},  all stars host solar-analog systems, and a 150-day bulge season is assumed. \label{fig:snow}}
\end{figure}

Figure \ref{fig:snow} shows the number of planetary detections as a function of $s$, for a total observing season of 150 days, and for various cadences. The detection is insensitive to the snowline index in the range $0.5\leq s\leq 1$, but there is a drop at $s>1$ such that, by $s=2$, the number of detections is lower by about 40\% compared to $s=0.5$.
Low values of $s$ keep the \textquoteleft Jupiters\textquoteright\ in our lens systems close to the Einstein radius, which scales with mass to the 1/2 power. Conversely, larger values of $s$ shift the giant planets away from the Einstein radius, where they are less likely to produce detectable anomalies (see Figure \ref{fig:planets}). The rocky planets that are then sometimes near the Einstein radius have masses that are too low for a probable detection.

As a concrete example, for the frequency of Solar-like systems estimated by \citet{Gould.2010.B}, of $f=1/6$, and assuming a snowline index of 1, the number of expected planet detections per season is $11.3 \pm 3.4$ in an optimal generation-II experiment. A four-year generation-II experiment can thus potentially produce a sizeable sample of order 50 microlensing detected planets. Conversely, for an assumed value of $s$, such an experiment will determine the abundance of snowline planets to a precision of $\sim15\%$, or even better if the actual planet frequency is higher.
Naturally, this result folds in our assumed scaled solar-analog model, with its particular range of planetary locations. This assumption leads to a predominance of gas and ice giants near the Einstein radius, though all types of planets can be near the Einstein radius due to projection effects (see Figure \ref{fig:planets}).

\subsection{Multiple planet detection and lensing planet identity}
\label{subsec:planet_identity}

\begin{figure*}
\begin{minipage}{\textwidth}
\center
\begin{tabular}{cc}
\includegraphics[width=0.5\textwidth]{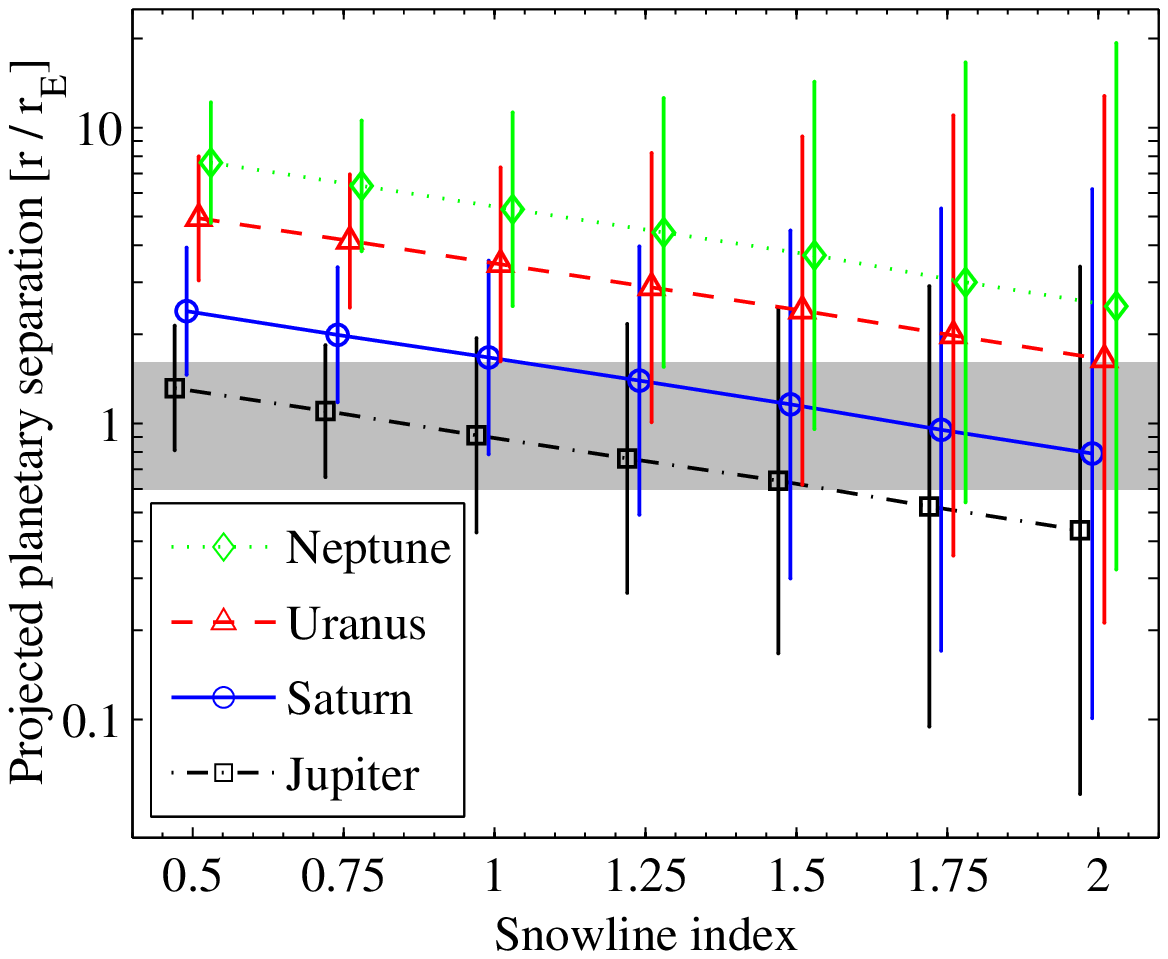} & \includegraphics[width=0.5\textwidth]{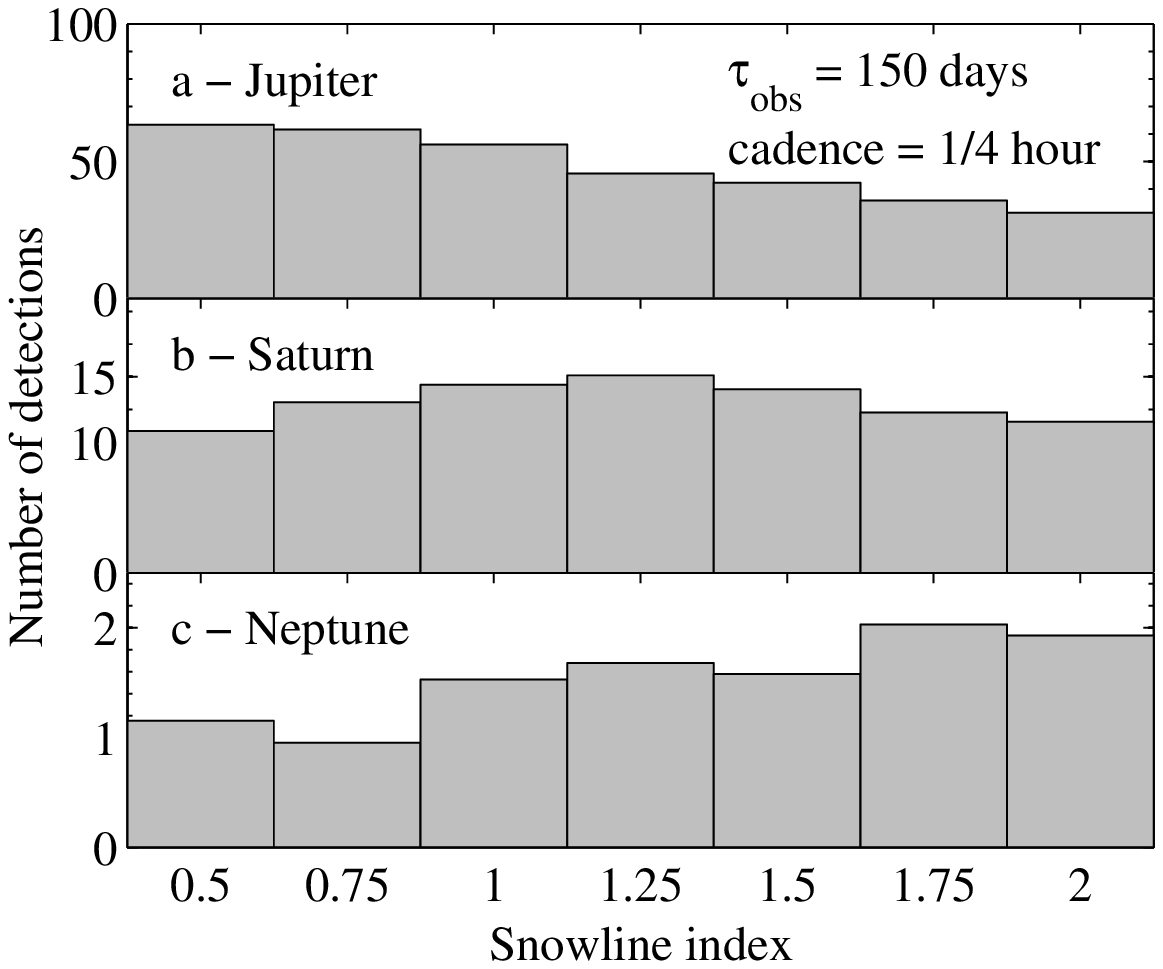}
\end{tabular}
\caption{ {\it Left}: Mean projected planetary separations of the simulated Jovian planets in units of the Einstein radius, $r_E=\theta_E \cdot D_L$, as a function of the snowline scaling index. Error bars indicates the RMS of each distribution. The gray area is the \textquoteleft\textquoteleft lensing zone\textquoteright\textquoteright\ , $0.6\leq r/r_E \leq 1.6$, where microlensing is most sensitive to planets. {\it Right}: Number of detections of \textquoteleft Jupiters\textquoteright\ , \textquoteleft Saturns\textquoteright\ and \textquoteleft Neptunes\textquoteright\ as a function of the snowline scaling index, for the optimal observing plan of 150-day season and minimal cadence of 15 minutes. \label{fig:planets}}
\end{minipage}
\end{figure*}

For the optimal observing plan of 150 days with 15-minute minimum cadence, we find that 83\% of the detections are due to deviations produced by a single planet, 16\% are due to two planets (i.e., two planets contribute to the detected anomalies in an event) and fewer than 1\% of the detections are caused by three of the planets in the system. Thus, if solar analogs are common, one out of every 4 or 5 planetary  detections could turn out to be a double-planet detection, as in OGLE-06-109L (\citealt{Gaudi.2008.A}; \citealt{Bennett.2010.A}).
In terms of the identity of the planets causing the anomalies, for a scaling index $s=2$, over 65\% of the planets are \textquoteleft Jupiters\textquoteright, 24\% are \textquoteleft Saturns\textquoteright, 4\% \textquoteleft Uranus\textquoteright, 4\% \textquoteleft Neptunes\textquoteright, and 1\% \textquoteleft Earths\textquoteright. The fraction of detected \textquoteleft Jupiters\textquoteright\ rises for lower snowline scaling indices, reaching 84\% for $s=0.5$.
Figure \ref{fig:planets} (left panel) shows the projected locations of the Jovian planets in units of the Einstein radius, $r_E=\theta_E \cdot D_L$, as a function of the snowline scaling index. Microlensing is most sensitive to planets in the so--called \textquoteleft\textquoteleft lensing zone\textquoteright\textquoteright, $0.6\leq r/r_E \leq 1.6$ (\citealt{Griest.1998.A}). The right panel of Figure \ref{fig:planets} shows the number of detections of Jupiters, Saturns and Neptunes as a function of the snowline scaling index. As seen in this figure, the lensing dominance of Jupiters among the detections is a consequence of the fact that, under our scaled solar-analog assumption, this planet (the most massive one) is never far from the Einstein radius. Nevertheless, medium-mass planets, such as Neptune, can also be detected with high cadence surveys, whenever they are projected near the lensing zone.


\section{SUMMARY}
\label{sec:summary}

In this work, we have studied various aspects of generation-II microlensing experiments -- the optimization of the observational parameters of such experiments, and their sensitivity to planets for different physical assumptions about the planetary systems. Our main results are the following:
\begin{enumerate}
\item Long duration experiments with continuous high cadence are vital. Experiments with low cadences of 3 hours will detect only about half as many planets as experiments with cadences of 15 minutes, and likewise experiments lasting one-half, versus a full, bulge season.
\item The current approach of using a network of 1-m-class telescopes for a generation-II experiment is adequate for tracking the low-magnification events discovered by these same telescopes. We have quantified, for our particular network and its parameters, the decrease of the planet detection efficiency with the decrease of the peak brightness of an event.
\item The abundance of planetary systems is degenerate with the snowline scaling index. Conversely, with constraints on planetary fraction from other planet-hunting techniques, microlensing surveys could discriminate among different planet formation scenarios that predict different snowline scalings.
\item The majority of events with anomalies will be due to single planet deviations (83\%), assuming Solar system distance and mass ratios, but the non-negligible remainder can reveal two planets.
\item For Solar-like systems, at least 65\% of the detections will be of \textquoteleft Jupiters\textquoteright, due to both the mass and the location near the Einstein radius of this planet.
\item The current generation-II experiment which continuously monitors 8 $\rm deg^2$ of the Galactic bulge with the highest microlensing event rate, and with a cadence of 15-30 minutes, will be able to determine, after four full seasons, the abundance of planets near the snowline to a precision of 10-30\%. This, assuming the fraction of Galactic stars hosting such planets is between 1/3 to 1/10, and a snowline scaling power-law index in the range 0.5 to 2.
\item The main contaminant to planet detection, binaries, can largely be filtered out already at the detection stage, based on the longer deviation timescales that they produce. Additional characterization and modeling of individual events in the timescale overlap region between planets and binaries will further lower the contamination fraction.
\end{enumerate}

A full generation-II experiment, of the type we have simulated here, is already in progress, our results show that first estimates of the frequency and nature of snowline-region planets will be possible soon. To improve the analysis, our future simulations will include also higher-order dynamical effects (parallax, orbital motion, xallarap), a distributions of source sizes, limb darkening, and a wider range of physical planetary system possibilities. Together with results from other planet-search techniques, a clearer picture should emerge of the properties of planetary systems and the physics of their formation.

\section*{Acknowledgments}

We thank S. Dong, A. Gould, A. Horesh, and E. Ofek for practical advice on lensing calculations, and A. Udalski for providing the OGLE-IV temporal sampling patterns. We thank OGLE, MOA, and PTF for their collaboration in the generation-II effort.



\bibliographystyle{mn2e}



\end{document}